# Spatial clustering of extreme annual precipitation in Uruguay

Florencia Santiñaque, Juan Kalemkerian, Madeleine Renom

Departamento de Métodos Cuantitativos, Facultad de Ciencias Económicas y Administración, Universidad de la República.

Centro de Matemática, Facultad de Ciencias, Universidad de la República.

Instituto de Física, Facultad de Ciencias, Universidad de la República.

## Abstract

The main objective of this work is to study the existence of spatial patterns maximum annual rainfall (through daily observations) within the territory of Uruguay and to show the application of two new statistical tools recently proposed. In the first stage, the distributions of maximum annual precipitation at each meteorological station will be studied. In the second stage, spatial clustering methods will be applied. To get the distribution of the maximum of each station, we have used a truncated Cramér-von Mises hypothesis test (the first statistical tool) and showed that it improves on the performance of the classic likelihood ratio test. It was found that in 18 study locations the distribution that best fits the data is of the Gumbel type, and for the other two, it is of the Fréchet type. Regarding the clustering methods, two methodologies were used, one of them was to perform clustering with the estimated parameters and the other was the PAM methodology using the F-madogram as distance, highlighting the homogeneity throughout the Uruguayan territory. Another novelty of this work (the second statistical tool) consists in including, as a complement to the clustering, the recently proposed independence test based on recurrence rates.

## 1. Introduction

The exploration and analysis of extreme meteorological data has been increasing due to the growth in climate variability. Various academic studies have focused on the use of extreme value theory in order to obtain conclusions in this regard. For example Portugués et al. (2008) studied the characteristics of extreme rainfall in La Rioja, Spain, analysing both the intensity (mm annual maximum daily) as well as the accumulation of rainfall as a consequence of the persistence of rain, over a certain period of time. Cartographies were produced reflecting the maximum intensity, magnitude and expected duration. In Hernández et al. (2011), extreme rainfall in Venezuela was studied, adjusting GEV (Generalized Extreme Value distributions, that is a family of distributions that includes Gumbel, Weibull and Fréchet distributions and that will be defined in Subsection 2.2) models from an estimation of the parameters, using Bayesian methods. The results showed that the Gumbel and Fréchet models are the most appropriate to represent the annual maxima in the studied locations. However, in locations with arid or very humid mesoclimates, the Weibull model is more appropriate. Other types of climatic variables have also been analysed using this method, among them Blanco et al. (2014), in which not only is the maximum rainfall studied, but there is also an analysis of the trend of the extreme temperatures in the State Durango, Mexico. Another climatic variable analysed using this type of method is the wind, e.g., in Fernández et al. (2016), which analyses the extreme speed of said phenomenon in Cuba, since obtaining this type of estimate is of the utmost importance, for example for structural design. Other works have combined the theory of extreme values with other types of methods such as copulas, clustering, and others (Moreno, 2013; Bernard et al., 2013; Bechler et al., 2015a;b). In Vannitsem et al. (2017), the maximum rainfall in Belgium was studied also using extreme value theory and incorporating information regarding spatial dependence. This work concluded that the degree of dependence on extreme rainfall in that country varies greatly according to three factors: the distance between two seasons, the season (summer or winter), and the duration of the accumulation of precipitation (per hour, day, month etc.). Rusticucci et al. (2010) studied such extreme events in South America. The performance of eight coupled global climate models (IPCC AR4) was studied in the simulation of the annual indices of extreme temperature climatic events and precipitation in South America. Two extreme temperature indices and three extreme precipitation indices were compared, based on information from meteorological stations from 1961–2000. Tencer et al. (2012) studied the interdecadal variability observed in the distribution of temperature events that exceed certain threshold, at five meteorological stations of Argentina, 1941–2000 period, by applying extreme value theory. The results showed a decrease in the intensity of extreme warm events over the study period, together with an increase in their frequency of occurrence during the last 20 years of the 20th century. The extremes of cold also show a decrease in intensity. However, changes in their frequency are not as consistent between the different stations studied. In Uruguay, however, there has been little study of extreme meteorological or climatic phenomena. Durañona (2015) studied strong winds and considered that the UNIT 50-84 (Uruguayan Institute of Technical Standards) standard should be reviewed and updated. The results obtained highlight, for example, that the geographical behavior of strong winds differs from those indicated on the national extreme winds map given by the UNIT 50-84 wind standard. Additionally, results evidenced that that the distribution of extreme winds averaged over 10 min for Montevideo can be properly modeled by a Gumbel distribution, while the UNIT 50-84 proposes Fréchet distribution for gusts of wind.

## 2. Material and Methods

In Subsection 2.1 we describe the objective of the work and the data set. In subsections 2.2 to 2.5 we include the methodology.

### 2.1 Data set and objective

This research was carried out with the aim of modeling the annual extreme rainfall accumulated over 24 hours (daily) in Uruguay as well as investigating the existence of spatial patterns in this phenomenon. There was a daily database of rainfall for the period 1981 to 2013 at 19 meteorological stations and 1 pluviometric station. In this framework, two objectives were set: 1. Study the distribution of extreme values of rainfall at each of the weather stations. The theory of extreme values provides a theoretical model to represent the behavior of the maxima recorded at different locations. 2. Identify the spatial patterns of extreme rainfall in Uruguay. In order to do this, spatial clustering methods have been used. The data set is the maximum annual rainfall from January 1981 to December 2013 at 20 weather stations located throughout Uruguay. This yields 33 observations for each one of the 20 weather stations. Figure 1 shows their locations, and the extreme rainfall boxplot for each one is given in Figure 2. It can be observed that the locations Artigas, Bella Unión, Colonia, Rocha, Salto, Treinta y Tres, and Young are the ones that registered annual extreme rainfall above 200 mm in different years. In 1997 there occurred important records in Bella Unión and in 1998 in Salto and Treinta y Tres. But Salto also stands out as the location that had the highest inter-annual variability in the behavior of this phenomenon. Melo stands out as the station with less inter-annual variability: the annual extreme rainfall did not exceed 150 mm in either year of the period under study.

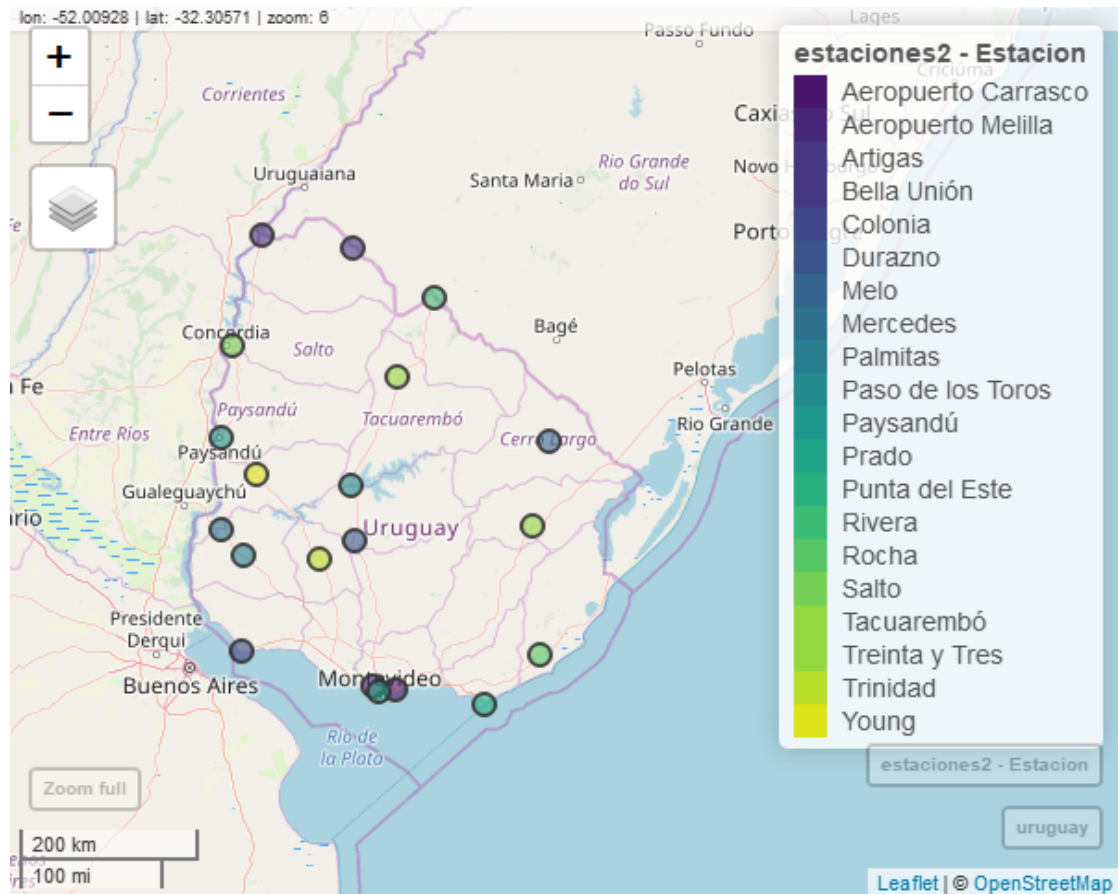

**Figure 1.** Map of Uruguay with the 20 weather stations analized in this work.

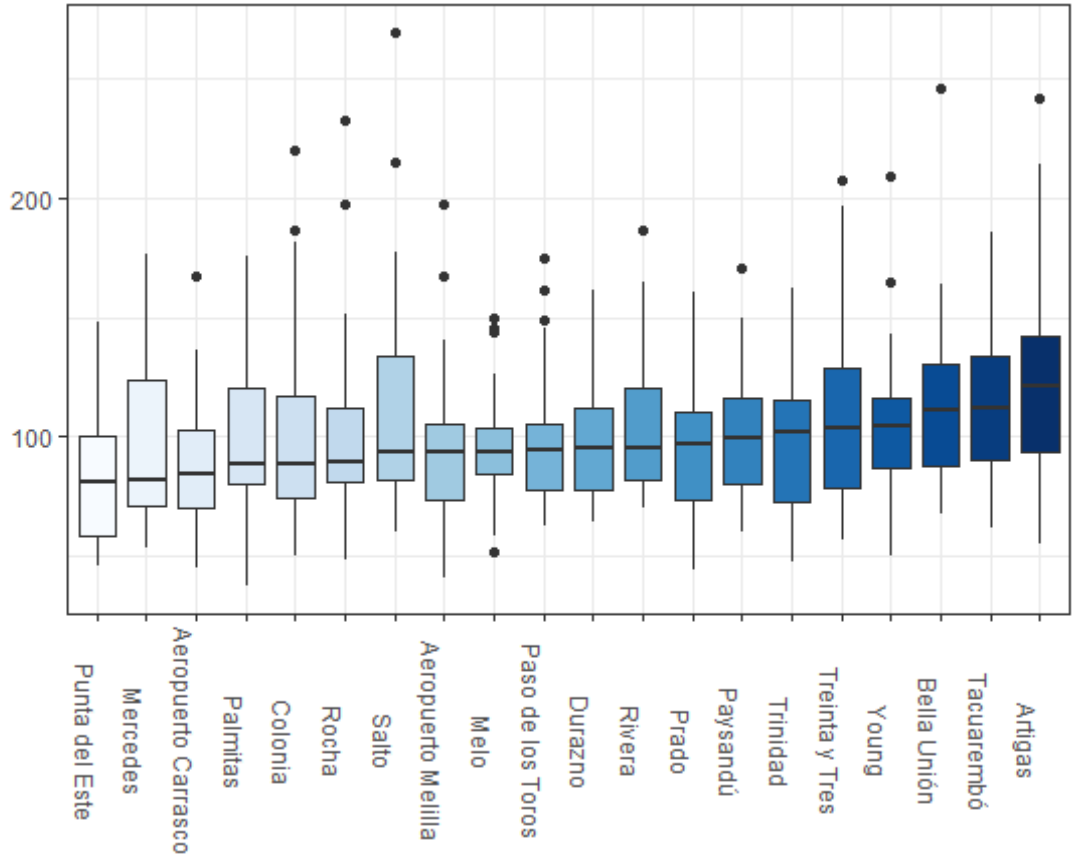

**Figure 2.** Maximum rainfall boxplot for each weather station.

## 2.2 Estimation of the distribution at each weather station

The literature evidences a profound development regarding the theory of extreme values (Resnick, 2007; de Haan and Ferreira, 2007; de Haan, 1978; Davison et al., 2012), and studies related to spatial statistics can also be found related to spatial statistics (Gaetan and Guyon, 2010), among others. According with the extreme value theory, the Fisher & Tippett theorem (Fisher and Tippett, 1928) formalized by Gnedenko (1948), said that if the sample size is large enough, then it is possible that the distribution of the maximum can be approximated by a Fréchet, Gumbel or Weibull family of distributions defined by: $H_1(x;\mu,\sigma) = e^{-e^{(\mu-x)/\sigma}}$ for $\sigma > 0$ (Gumbel), $H_2(x;\mu,\sigma,\xi) = e^{-\left(\frac{x-\mu}{\sigma}\right)^{-1/\xi}}$ where $x > \mu, \sigma, \xi > 0$ (Fréchet) and $H_3(x;\mu,\sigma,\xi) = e^{-\left(\frac{\mu-x}{\sigma}\right)^{-1/\xi}}$ where $x < \mu, \sigma > 0\ \xi < 0$ (Weibull). These three types of functions can be included in the following expression: $H(x;\mu,\sigma,\xi) = e^{\left(-1+\frac{\xi(x-\mu)}{\sigma}\right)^{-1/\xi}}$ where $\sigma > 0$ and $x > \mu - \sigma/\xi$ for $\xi > 0$ or $x < \mu - \sigma/\xi$ for $\xi < 0$. The distribution $H$ is Fréchet when $\xi > 0$, Weibull when $\xi < 0$, and if $\xi \to 0$, then

$H$ becomes a Gumbel distribution. $H$ is called the Generalized Extreme Value distribution (GEV). Given $p \in (0,1)$ we can find $z_p$ such that $H(z_p) = 1 - p$. It is known that if we take a return period $t = 1/p$, then $z_p$ is the return level. The parameters μ, σ and ξ (location, scale and shape respectively), were estimated by three methods: classical maximum likelihood, and two methods designed (and widely used) for extreme value statistics: profile likelihood and the method of weighted moments. The method called weighted moments was proposed by Greenwood et al. (1979). An advantage of the profile maximum likelihood method is that it allows a non-symetric confidence interval. In several cases, for extremes it can be more reasonable to have non-symmetric intervals.

**2.3 Model diagnosis.**

Once we have estimated the parameters, as the second step, we will examine a goodness of fit test for the distribution of each station. $H_0 : X^{(i)} \sim$ Gumbel(μ, σ) vs $H_1 : H_0$ does not hold, where $X^{(i)}$ is the yearly maximum of rainfall at station *i*. If $H_0$ is rejected, then we perform a test for the Fréchet distribution (if the estimation of ξ is positive) or Weibull (if the estimation of ξ is negative). All R packages concerning extreme value statistics include the likelihood ratio test that assumes that the distribution of the observed sample obeys a GEV distribution. In our case, we do not have a large sample size, also we prefer to use a test that does not assume any distribution previously. In addition, the p-value for the likelihood ratio test is calculated from the asymptotic distribution, which can lead to error given our moderate sample size (33). To adjust the distribution of each station, we used a truncated Cramér–von Mises test for the Gumbel distribution. We adapt the idea proposed in Kalemkerian (2019) to the Gumbel distribution. The details of this adaptation can be found in Santiñaque (2020) (pages 24 and 25).

To add more validity to the results obtained by the goodness of fit test, in each station, we have made diagnostic plots that compare the empirical values vs the adjusted values in the four ways that are listed below. 1: PP plots compare the theoretical cummulative probabilities with the empirical cummulative probabilities. 2: QQ plots compare the empirical quantile function with the adjusted quantile function. 3: Empirical density vs adjusted density. 4: Return period–return level plot. In the PP plot and QQ plot, if the points are close to the diagonal, the adjusted distribution works well. In the return period/return level, if the points are near the straight line, the Gumbel distribution is suitable, whereas if the points are near the dashed curve above the straight line, the Fréchet distribution is suitable. But if the points are near the dashed curve below the straight line, the Weibull distribution is appropriate, see for example Coles et al., (2001).

## 2.4 Spatial Clustering

The main objective of this section is to investigate whether there are groups of stations with similar behavior in terms of maximum annual rainfall, and if so, how many groups there are and where they are located geographically. Of course, we can take one, two or three groups according with the type of distribution that has been adjusted in each station. However we are interested in applying standard clustering methods and comparing that with a method designed and used for extreme events. On the one hand, we have performed clustering of the estimated parameters. We applied a hierarchical Ward method with Euclidean distance. Also, we applied a non-hierarchical PAM method (Partitioning around medoids) proposed by Kaufman and Rousseeuw (1990). Unlike the *K*-means method, where each cluster is represented by its mean, in the PAM method each cluster is represented by a particular observation in it (medoid). Thus, when the observations are maximal, the medoid of each cluster remains a maximum, which does not happen in *K*-means. For this reason, the PAM method to obtain clusters looks more reasonable than the *K*-means method. Also, the PAM method is more robust than the *K*-means method (Izenman, 2008). To obtain the optimal number of groups we have used the Silhouette coeficient. The graphical tool called Silhouette was proposed by Rousseeuw (1986). On the other hand, we have performed the clustering method proposed in Bernard et al. (2013) where it was applied with good results to detect spatial dependencies between heavy rainfall events in France. The method consists of applying PAM clustering using the distance defined from the F-madogram, proposed by Cooley et al. (2006) and Naveau et al. (2009). A good explanation of this method can be found in Bernard et al, (2013).

## 2.5 Independence test based on recurrence rates between pairs of stations

As a complement to what was done in the clustering subsection, we consider the independence test between two variables proposed by Kalemkerian and Fernández (2020a). The test can be summarized as follows. Given a sample $(X_1, Y_1), (X_2, Y_2), ..., (X_n, Y_n)$ of $(X, Y)$ where $X \in S_X$ and $Y \in S_Y$, where $S_X$ and $S_Y$ are metric spaces, we want to test $H_0: X$ and $Y$ are independent vs $H_1: H_0$ does not hold. The test is based on a function that measures the difference between the joint recurrence rates between X and Y, and the product between the marginal recurrence rates X and Y. The implementation of the test and its theoretical properties can be found in Kalemkerian and Fernández (2020a). In Kalemkerian and Fernández (2020b) other climatological applications of the test can be found as well as its good performance under a wide spectrum of alternatives.

## 3. Results and Discussions

### 3.1 Parameter Estimation

The parameter estimates obtained by the three methods were similar. In Table 1, in columns 2 to 5, we show the maximum likelihood estimated values of μ, σ and ξ, and a 95% confidence interval for ξ from the profile maximum likelihood. Except for Mercedes, all the confidence intervals for ξ contains the value of zero, which suggest that several of the stations considered in this work can be modeled by a Gumbel distribution. We refine this fact using the goodness of fit test. Column 6 of Table 1 shows the *p*-values for the goodness of fit test for the Gumbel distribution for each station. We conclude that at the 5% level, there is empirical evidence that

Rocha and Mercedes do not have a Gumbel distribution. For Rocha and Mercedes, once the test rejected the hypothesis of a Gumbel distribution, we made the same adaptation to test the goodness of fit for the Fréchet distribution, and the *p*-values were 0.10 and 0.33, respectively. In summary, the truncated test of fit goodness is that Rocha and Mercedes have a Fréchet distribution, and the other stations have a Gumbel distribution.

| Station | $\hat{\mu}$ | $\hat{\sigma}$ | $\hat{\xi}$ | 95% C. I. for $\xi$ | *p*-value (Gumbel) |
|---|---|---|---|---|---|
| Punta del Este | 70.25 | 22.30 | -0.04 | (-0.35,0.35) | 0.56 |
| Aeropuerto Carrasco | 75.56 | 21.36 | 0.01 | (-0.28,0.23) | 1.00 |
| Mercedes | 78.13 | 24.54 | 0.17 | (0.02,0.76) | 0.03 |
| Colonia | 80.41 | 27.7 | 0.15 | (-0.14,0.53) | 0.46 |
| Aeropuerto Melilla | 81.13 | 28.52 | -0.08 | (-0.29,0.15) | 0.32 |
| Rocha | 81.99 | 18.81 | 0.27 | (-0.09,0.33) | 0.03 |
| Prado | 82.32 | 27.77 | -0.19 | (-0.46,0.05) | 0.43 |
| Paso de los Toros | 84.66 | 19.94 | 0.10 | (-0.19,0.48) | 0.54 |
| Palmitas | 84.78 | 29.16 | -0.11 | (-0.41,0.10) | 0.28 |
| Melo | 86.49 | 21.49 | -0.12 | (-0.40,0.10) | 0.45 |
| Durazno | 86.78 | 21.07 | -0.05 | (-0.35,0.33) | 0.60 |
| Trinidad | 87.91 | 29.05 | -0.20 | (-0.47,0.07) | 0.19 |
| Paysandú | 88.04 | 24.87 | -0.08 | (-0.39,0.23) | 0.23 |
| Salto | 89.05 | 25.16 | 0.25 | (-0.03,0.61) | 0.16 |
| Rivera | 89.50 | 19.76 | 0.17 | (-0.08,0.09) | 0.36 |
| Young | 89.97 | 24.60 | -0.04 | (-0.23,0.16) | 0.56 |
| Treinta y tres | 90.92 | 31.77 | 0.09 | (-0.17,0.58) | 0.26 |
| Bella Unión | 97.67 | 26.96 | 0.03 | (-0.21,0.34) | 0.89 |
| Tacuarembó | 99.53 | 30.22 | -0.12 | (-0.47,0.19) | 0.64 |
| Artigas | 103.32 | 39.96 | -0.04 | (-0.35,0.30) | 0.89 |

**Table 1.** Maximum likelihood estimated values of µ, σ and ξ, and a 95% confidence interval for ξ from the profile maximum likelihood. Column 6 shows the *p*-values for the truncated Cramér-von Mises test for the Gumbel distribution.

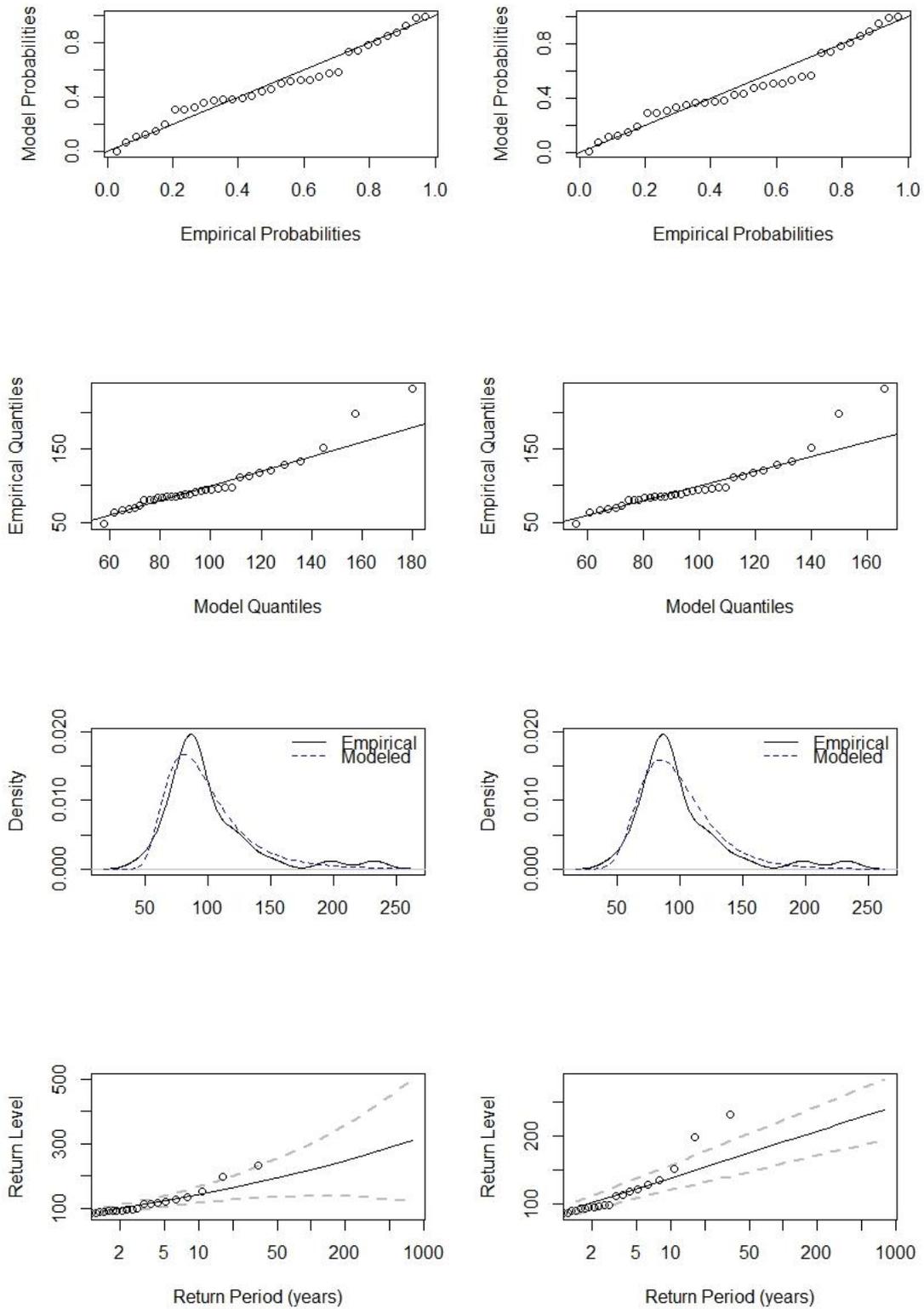

**Figure 3.** Diagnostic plot for Rocha adjusted distribution from Fréchet model (left) vs Gumbel model (right).

### 3.2 Model diagnosis

The diagnostic plots showed that the adjusted distribution works well for each one of the 20 stations. For example, in Figure 3 we show the four diagnostic plots for Rocha station, using Fréchet and Gumbel distribution. It is important to note that we have obtained improvements using the truncated Cramér-von Mises test instead of the likelihood test. For example, Rocha station according to the likelihood ratio test is adjusted by Gumbel distribution (instead Fréchet according with the truncated Cramér-von Mises test). Figure 3 shows a better fit to Fréchet than to Gumbel distribution in the qq plots and the return level plots.

### 3.3 Clusters of the estimated parameters

Applying a hierarchical Ward method with Euclidean distance, the adjusted $R^2$ defined as $\frac{R^2/(K-1)}{(1-R^2)/(n-K)}$ where $K$ is the number of groups, has a maximum for $K = 2$ groups. In Figure 4 we show the dendrogram for $K = 2$ and $K=3$ groups.

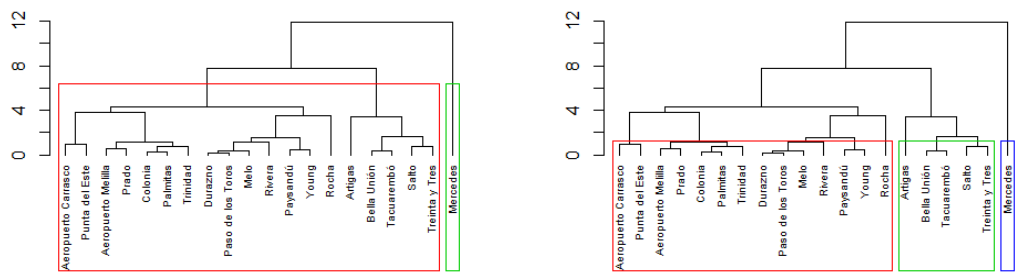

**Figure 4.** Dendrogram for $K=2$ and $K=3$ groups from Ward method with Euclidean distance.

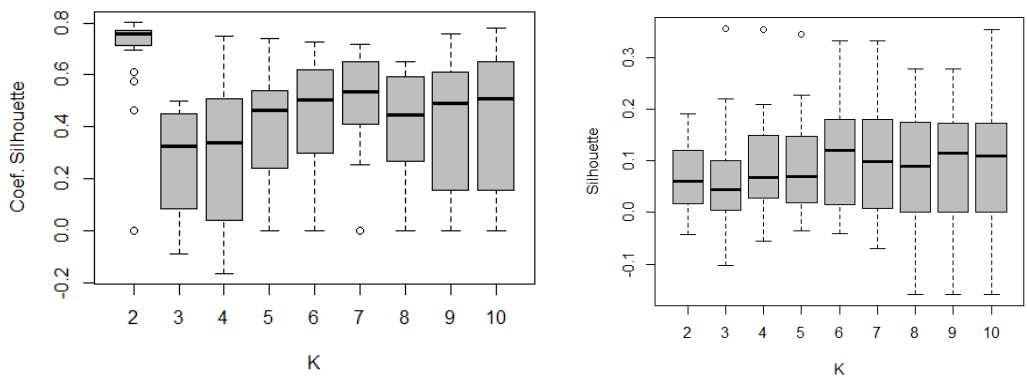

**Figure 5.** Silhouette coefficient for different values of $K$ using the estimated parameters (left) and $F$-madogram as a distance (right).

Figure 4 and Figure 5 (left) show that the Silhouette coefficient clearly suggests that there are two groups: one is Mercedes by itself, and the other one consists of the remaining 19 stations. Observe that using the PAM methodology with F-madogram the values of the Silhouette coefficient are poor for all values of *K* considered. In summary, using the estimated parameters, the existence of two different groups is clear. Taking into account that one of these groups is formed by only one station (Mercedes), we can deduce that there are no substantial differences between the yearly maximum rainfall throughout the geographical area studied. On the other hand, considering the PAM method with the F -madogram, the Silhouette coefficient does not detect a clear separation into groups, which reinforces the conclusion that the behavior of the yearly maximum rainfall is homogeneous throughout Uruguay.

In Figure 6 we show the geographical distribution of the different stations, separating them into 2 (the optimum number of groups according with Silhouette criterion), 3 and 4 groups.

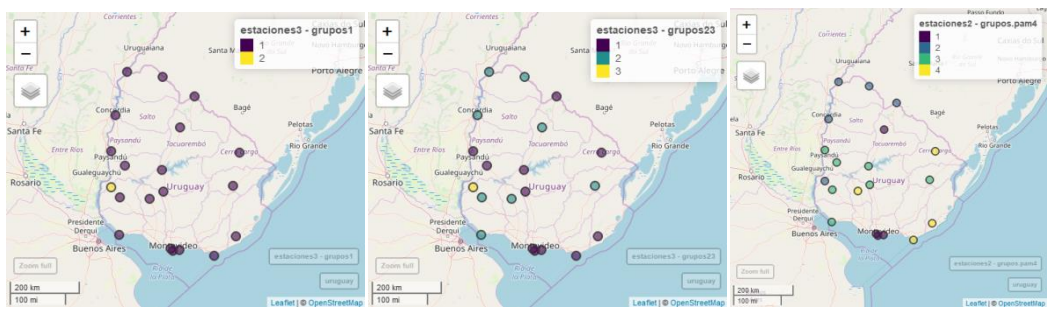

**Figure 6.** Geographical distribution separating into 2, 3 and 4 clusters.

### 3.4 Independence test based on recurrence rates between Mercedes and the other stations.

In Table 2 we show the p-values for the independence test based on the recurrence rates between Mercedes station and each one of the remaining stations. From Table 2 we can conclude that if we work at the 5% level of significance, Mercedes did not reject the null hypothesis of independence of each one of the other stations. It can also be shown that Rivera did not reject the null hypothesis of independence, neither with Artigas nor with Bella Unión. Also for the stations in the metropolitan area (Melilla, Prado and Carrasco Airport), the null hypothesis of independence is not rejected. Also, for Rocha and Punta del Este, the null hypothesis of independence is not rejected. If the results of the independence test were taken as a grouping criterion, it can be seen that several of the results obtained by the previous clustering methods are reinforced by the results obtained by this test.

### 4. Conclusions

We have studied the existence of spatial patterns of daily maximum annual rainfall within Uruguay using two clustering techniques. One technique was to adjust a GEV distribution for each one of the 20 stations, and then make clusters from their sets of estimated parameters. With this, the Silhouette coefficient clearly suggests two groups, but one of them consists exclusively of Mercedes station. In other technique, we have used the PAM methodology using the F-madogram as the distance. With this, the Silhouette coefficient does not suggest grouping stations. In addition, we have introduced two statistical techniques. First, to adjust the distribution for each station, we have adapted a truncated Cramér–von Mises test of normality to test a GEV distribution; this test achieved a better result than the classical likelihood ratio test. Second, we have applied the recently proposed independence test based on recurrence rates, and found that this test can be used as a complement to a clustering analysis to see if each station belonging to one group is statistically independent of each station corresponding to another group. As a final conclusion of the entire study, we can conclude that throughout the Uruguayan territory, the behavior of the maximum annual rainfall is homogeneous, with the particularity of Mercedes whose observations were independent from the rest of stations (according to the independence test based on recurrence rates). Also, Mercedes is the only member of a group when clusters of two groups are made.

| Carrasco | Melilla | Artigas | Bella Unión | Colonia |
|---|---|---|---|---|
| 0.61 (0.918) | 0.96 (0.948) | 0.31 (0.839) | 0.18 (0.722) | 0.58 (0.914) |
| Durazno | Melo | Palmitas | Paso de los Toros | Paysandú |
| 0.05 (0.000) | 0.63 (0.921) | 0.61 (0.918) | 0.13 (0.615) | 0.79 (0.937) |
| Prado | Punta del Este | Rocha | Rivera | Salto |
| 0.67 (0.925) | 0.62 (0.919) | 0.15 (0.667) | 0.69 (0.927) | 0.69 (0.927) |
| Tacuarembó | Treinta y Tres | Trinidad | Young | |
| 0.71 (0.929) | 0.05 (0.000) | 0.19 (0.737) | 0.55 (0.909) | |

**Table 2.** *p*-values and the percentage of exceedance with respect to the significance level of 5% (in parentheses) for the independence test based on recurrence rates between Mercedes and the remaining stations.